\documentclass[useAMS,usenatbib]{mn2e}

\usepackage{graphicx}
\usepackage{url}
\usepackage{float}
\usepackage{color}
\usepackage{soul}

\usepackage{txfonts}

\title[Spectral analysis of BD+30$^{\circ}$623, the CS of NGC 1514]{Spectral analysis of BD+30$^{\circ}$623, the
  peculiar binary central star of the planetary nebula NGC 1514\thanks{Based on 
observations obtained at the German-Spanish Astronomical Center, Calar Alto,
jointly operated by the Max-Planck-Institut f\"ur 
Astronomie (Heidelberg) and the Instituto de Astrof\'{\i}sica de
Andaluc\'{\i}a (CSIC)}}

\author[A. Aller et al.]{A. Aller$^{1,2,3}$\thanks{E-mail:
    alba.aller@cab.inta-csic.es}, B. Montesinos$^{1}$, L. F. Miranda$^{4,3}$, E. Solano$^{1,2}$, A. Ulla$^{3}$\\
$^{1}$Departamento de Astrof\'{\i}sica, Centro de Astrobiolog\'{\i}a
  (INTA-CSIC), PO Box 78, E-28691 Villanueva de la Ca\~nada (Madrid),
  Spain\\
 $^{2}$ Spanish Virtual Observatory, PO Box 78, E-28691 Villanueva de la Ca\~nada (Madrid),
  Spain\\
$^{3}$Departamento de F\'isica Aplicada, Universidade de Vigo, Campus
  Lagoas-Marcosende s/n, E-36310 Vigo, Spain\\
   $^{4}$Instituto de Astrof\'{\i}sica de Andaluc\'{\i}a - CSIC, C/ Glorieta de la Astronom\'{i}a s/n, E-18008 Granada, Spain
}
\begin{document}

\date{Accepted 1988 December 15. Received 1988 December 14; in
  original form 1988 October 11}

\pagerange{\pageref{firstpage}--\pageref{lastpage}} \pubyear{2002}

\maketitle

\label{firstpage}

\begin{abstract}
NGC 1514 is a complex planetary nebula with a peculiar binary central star (BD+30$^{\circ}$623)
 consisting of a cool star and a hot companion. To date, the parameters of the two stars have not been firmly stablished. 
We present a detailed spectral analysis of BD+30$^{\circ}$623 based on intermediate-resolution CAFOS
optical spectra and IUE ultraviolet spectra with the goal of deriving the parameters of the two stars. 
For this purpose, we used an extensive composite grid of
Kurucz and T\"{u}bingen NLTE Model-Atmosphere spectra. From the
fitting procedure, in terms of the minimum $\chi^{2}$ method, the best models obtained 
correspond to an Horizontal-Branch A0 star with $T_{\rm eff}$ = 9850$\pm$150\,K, $\log g$
= 3.50$\pm$0.25, and a hot companion with $T_{\rm eff}$ between 80000\,K and 95000\,K and a $\log
g$ $\simeq$ 5.5. To our knowledge, this is the first time that the 
parameters of both stars have been determined accurately through a detailed spectroscopic analysis.

\end{abstract}


\section{INTRODUCTION}
\label{Section:introduction}

Central stars of planetary nebulae (CSPNe) are evolved remnants of low- and
intermediate-mass stars ($0.8\le M/{{\rm M}_{\odot }}\le 8$) and immediate precursors 
of white dwarf stars. Although they are key in the
 formation and evolution
 of their associated PNe, many
aspects of CSPNe remain unknown by now. Among these, the binarity and
its relationship with the properties of the nebulae is one of the most
puzzling issues. According to Frew \& Parker (2010) there are more
than 3000 known galactic PNe. However, there is only spectroscopic information for 
$\sim$13\% of their CSs (Weidmann \& Gamen, 2011). Recent works show
that just $\sim$ 40 of the exciting stars studied have been catalogued
as binary CSs (de Marco et al. 2013). Therefore, there is
still very little information concerning binarity in CSPNe which
prevent us from constraining the formation and properties of these
systems. In this context, many studies are being carried out with the aim of clarifying the role
 of binary stars in the formation of PNe (see, e.g., Miszalski et al. 2009).
 
 There are many types of binary CSs but ones of special interest are those named 
 {\it peculiar central stars} (Lutz 1977). This class refers to cool (spectral type A through K)
  CSs that are not hot enough to ionize their associated nebulae. Lutz suggested that these
stars may belong to binary systems, where a hot (and faint)
component would be the responsible for the photoionization while the cooler
 (and brighter) star would account for the absorption
spectrum. Many works can be found regarding these stars (see, e.g, M\'endez 1978, 
de Marco 2009, Pereira et al. 2010). However, the complexity of these systems makes 
their analyses difficult because the determination of the stellar parameters is laborious.

BD+30$^{\circ}$623 ($\alpha$ $\! = \!$ 04$^ {\rm h}$\,09$^ {\rm m}$\,16$\fs$9,
$\delta$ $\! = \!$ +30$^{\circ}$\,46$'$\,33$''$, equinox 2000.0; {\it $\ell$} $\! = \!$
165$\fdg$53, {\it b} $\! = \!$ -15$\fdg$2), the exciting CS of NGC 1514, is one of
 these peculiar binary stars. It has been studied extensively but the physical 
 parameters of the two components have not been determined accurately to date. It
was initially classified as a single star and several spectral types were proposed in the literature ranging
 from B8 (Seares \& Hubble 1920), to B9 (McLaughlin 1942) or A0 (Chopinet 1963). In
contrast, Payne (1930) proposed an O8
classification.

Because of its peculiar spectrum, Kohoutek (1967) proposed, for the first
time, the double star hypothesis for BD+30$^{\circ}$623 based on
photoelectric photometry. He reported the existence of a fainter and
hotter companion and described the pair as A0 III + blue subdwarf
(sdO) with effective
temperatures ($T_{\rm eff}$) of 10800\,K for the A-star and 60000\,K for the hot
star, and radii of 4.1${\rm R}_{\odot }$ and 0.45 ${\rm R}_{\odot }$,
respectively. Later, Kohoutek \& Hekela (1967) confirmed these values
based on a spectral analysis of the CS within the interval
$\lambda\lambda$ 3650-5000\,$\AA$. In contrast, Greenstein (1972)
reported that the ultraviolet luminosity of the sdO required a higher
$T_{\rm eff}$ (100000\,K), and that the
cool star was, in fact, an Horizontal-Branch A star (A3- A5) with
$T_{\rm eff}$ $\le$ 10000\,K. Further evidence of the binary nature of this system
 was provided by
the International Ultraviolet Explorer (IUE) observations obtained by
Seaton (1980). With these UV spectra and broad-band photometry from
the Dutch ANS satellite, he constructed a model and described the
system as an A0-A3 III star with $T_{\rm eff }$ $\sim$
9000\,K, and $\log g$= 3.0, and a hot star with $T_{\rm eff}$ $\ge$
60000\,K. To do this, he used a line blanketed model for the cool star
and a blackbody for the hot companion. Besides, weak P Cygni profiles
were detected. Later, Feibelman (1997) also used IUE spectra to report a
considerable variability of the UV flux (by a factor of two) but he
was not able to explain it. In strong contrast with previous determinations, 
Grewing \& Neri (1990) used different methods to estimate $T_{\rm eff}$
 of 27000, 28000, and 38000\,K for the hot component. These values are too low to account 
 for the observed He\,{\sc ii} emission from the nebula (Kaler 1976), which requires $T_{\rm eff}$$>$ 60000\,K (see Pottasch 1984).
In a most recent work, Taranova \& Shenavrin (2007)
 proposed a spectral type B(3-7) main-sequence for the cool star based on infrared
photometry. In these circumstances, it is clear that a new and detailed spectral analysis is necessary
 in order to determine the properties of this pair.

 In this framework, 
 we present an innovative spectral analysis of BD+30$^{\circ}$623 
by means of IUE ultraviolet and intermediate-resolution
optical spectra. For this purpose, grids of synthetic spectra for the
cool and hot stars were used and combined making use of the latest and state-of-the-art
synthetic model-atmospheres. As far as we know, this is the first time that such analysis
is done for one of these peculiar binary CSs.

\section{OBSERVATIONS AND RESULTS}
\label{Section:observations}

\subsection{Intermediate-resolution optical spectra}

\begin{figure}
\includegraphics[width=0.5\textwidth]{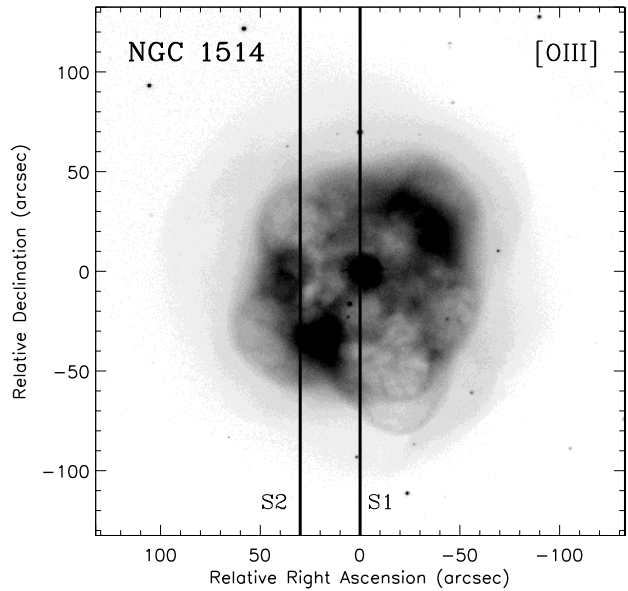}
  \vspace*{1pt}
  \caption{grey-scale reproduction of the [O\,{\sc iii}] image of
    NGC 1514. grey levels are linear. The two vertical lines S1 and S2 
    mark the slit positions used for intermediate-resolution, long-slit spectroscopy (see text). }
\label{Figure:CAFOSimage}
\end{figure}

Intermediate-resolution, long-slit spectra were obtained on 2011
January 16 with the Calar Alto Faint Object Spectrograph (CAFOS) at
the 2.2-m telescope at Calar Alto Observatory (Almer\'{i}a,
Spain). A SITe 2k$\times$2k--CCD was used as detector. Gratings B-100
and R-100 were used to cover the 3200--6200 $\AA$ and 5800--9600 $\AA$
spectral ranges, respectively, both at a dispersion of $\simeq$
2\,$\AA$\,pixel$^{-1}$. The slit width was 2\,arcsec and the spectra were
obtained at two slit positions, both oriented north -- south: one
(denoted S1) with the slit centered on the CS and with an
exposure time of 100\,s for each grism; and another (S2) displaced
$\sim$ 30\,arcsec eastern from BD+30$^{\circ}$623 and with an exposure
time of 1800\,s for each grism in order to cover the nebula. The
projections of the slits on the sky are plotted in
Fig.\,\ref{Figure:CAFOSimage}, which shows an [O\,{\sc iii}]
($\lambda_{\rm 0}$ = 5007 \AA, FWHM = 87 \AA) image of NGC 1514 also
obtained with CAFOS in imaging mode. Seeing was $\simeq$ 2\,arcsec
during the observations.

The spectra were reduced with standard routines for long-slit
spectroscopy within the {\sc iraf} and {\sc midas} packages. For the
absolute flux calibration, the spectrophotometric standards G191B2B
and Feige\,34 were observed the same night. We note
that the red part of the spectrum (above $\sim$ 6200 $\AA$) presents 
an unexpected behavior that we were unable to correct in the flux calibration process.
 Therefore, the wavelength range beyond 6200 $\AA$ has not been
taken into account for the analysis. However, the normalized
uncalibrated spectrum at these wavelengths, in particular around
H$\alpha$, is still usable to estimate the gravity of the cool
component (see Sect. \ref{Section:cool}). Finally, we note 
that no problems with light losses in the flux calibration are present, since 
the available photometry in the blue part of 
the spectrum is compatible with our flux calibration.

The optical spectrum of BD+30$^{\circ}$623 (Fig.\,2), obtained at S1,
 shows strong hydrogen absorption lines which are typical of A- or B-type stars. The
Ca\,{\sc ii} K $\lambda$3933 absorption line, also characteristic of
these stars, is present. The spectrum also
reveals other spectral features like He\,{\sc ii} $\lambda$4686 and
$\lambda$5412 which are not expected in a typical spectrum of an A- or
B-type star. They only
appear (specially He\,{\sc ii} $\lambda$4686) in stars with $T_{\rm eff}$ $\gtrsim$
40000\,K (Eisenstein et al. 2006).

\begin{figure*}
\includegraphics[width=0.9\textwidth]{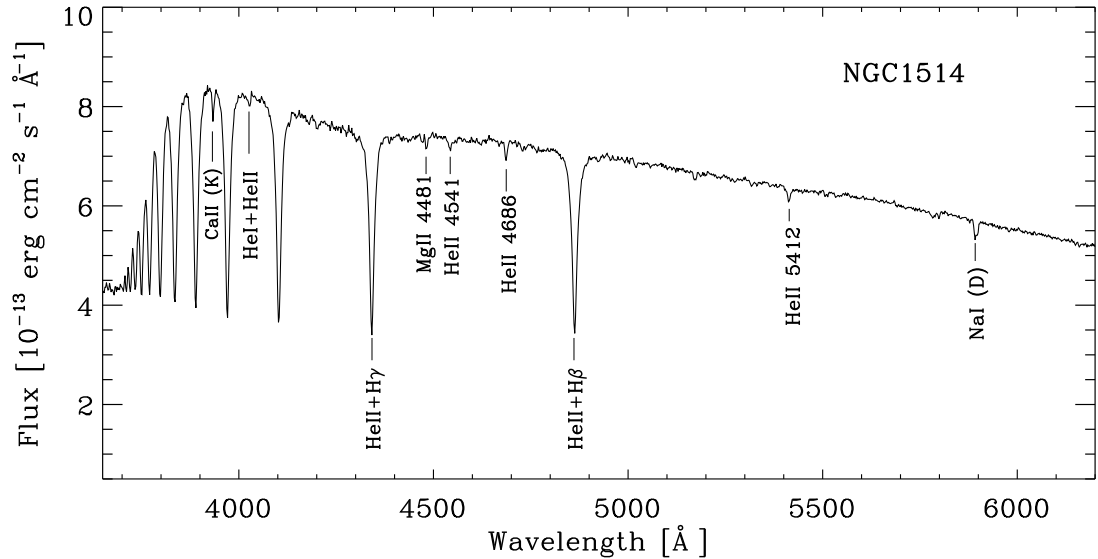}
  \vspace*{1pt}
  \caption{CAFOS CAHA blue spectrum of BD+30$^{\circ}$623
    in the range 3500--6200 $\AA$. Some helium and hydrogen absorption
    lines are indicated, as well as the Ca\,{\sc ii}\,$\lambda$3968,
    and Mg\,{\sc ii}\,$\lambda$4481. The Na\,{\sc i} doublet
    ($\lambda$5889-5895) is caused by interstellar absorption.}
\end{figure*}

On the other hand, the spectrum shows very weak features of neutral
helium. It is the case of He\,{\sc i} $\lambda$4026 (whose
contribution cannot be separated from that of He\,{\sc ii}
$\lambda$4025 at this resolution) and He\,{\sc i} $\lambda$4471 (that
seems to be also present although with a smaller contribution). These
neutral helium absorption lines may originate either from the cool
star or from the hot companion. Once the spectral analysis is concluded (see Sect. \ref{Section:analysis})
we will see that these neutral helium features come from the cool companion. 

We note that no nebular emission lines are observed in Fig.\,2 
 because of the very short exposure time of this spectrum (100\,s), 
 but they are clearly observed in the long-exposure spectrum obtained at S2 (Fig.\,1).
We used this spectrum (not shown here) to 
determine the logarithmic extinction coefficient $c$(H$\beta$). Because the
H$\alpha$ emission line flux is unreliable (see above), we used the integrated
H$\beta$ and H$\gamma$ fluxes observed at slit position S2 (see Fig.\,\ref{Figure:CAFOSimage}) to 
derive $c$(H$\beta$) $\simeq$ 0.97, assuming Case B recombination
  ($T_{\rm e}$=10$^{4}$\,K, $N_{\rm e}$=10$^{4}$\,cm$^{-3}$) and a
  theoretical H$\gamma$/H$\beta$ ratio of  0.466 (Osterbrock
  1989). If we derive the colour excess from the relationship proposed by Seaton
  (1979) $c$(H$\beta$) = 1.47 $E(B\!-\!V)$, we obtain that
  $E(B\!-\!V)$= 0.66, in agreement with the value derived from the
  analysis of the CS (see Sect. \ref{Section:extinction}). We also note that a slightly lower value for the logarithmic
  extinction coefficient ($c$(H$\beta$) = 0.885) was derived by Muthu (2001) 
  from the observed H$\alpha$/H$\beta$ ratio. Finally, the detection of the He\,{\sc ii} $\lambda$4686 
  emission line allows us to set a lower limit of $T_{\rm eff}$ = 60000\,K for the hot companion (see above).

\subsection{Low-resolution ultraviolet spectra}
\label{Section:IUE_spectra}

Since the $T_{\rm eff}$ of the hot companion of
BD+30$^{\circ}$623 should be higher than 60000\,K,
observations in the ultraviolet range are crucial to achieve a good characterization of this star.
 For this purpose,  we
have retrieved spectra from the International Ultraviolet Explorer
(IUE) (Kondo et al. 1989) available in the IUE Newly Extracted Spectra
(INES\footnote{\tt{http://sdc.cab.inta-csic.es/ines/}})
System. INES provides spectra already calibrated in physical units.

We have retrieved the available low-dispersion ($\sim$ 6\,$\AA$) IUE
SWP (short wavelength) and LWP (long wavelength) spectra, which were
obtained from 1978 to 1989. The SWP and LWP cover the ranges 
1150-1975\,$\AA$ and 1910-3300\,$\AA$, respectively. 
Only the pairs of spectra which were obtained consecutively, i.e., the same
day were used. They are listed in Table\,\ref{Table:IUEspectra}. A mean of all
spectra was calculated for the spectral analysis. 
Unfortunately, the LWP spectra beyond
 $\sim$ 2400 \AA{} could not be used since most of them appear to
be saturated and/or with many bad pixels at those wavelengths.

\begin{figure}
\includegraphics[width=0.4\textwidth]{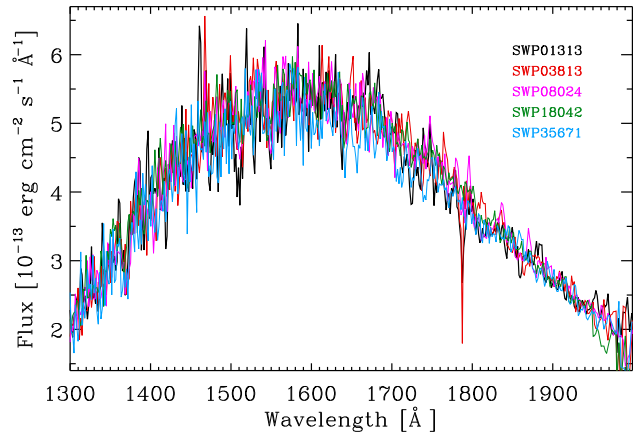}
  \vspace*{1pt}
  \caption{Plot of the five SWP spectra of BD+30$^{\circ}$623
    listed in Table\,\ref{Table:IUEspectra} and used for the spectral
    analysis. No variability is observed in the SWP spectra, in
    contrast with that reported by Feibelman (1997).}
\label{Figure:IUEspectra}
\end{figure}

Fig. \ref{Figure:IUEspectra} shows the SWP spectra listed in Table
\ref{Table:IUEspectra}. Surprisingly, we have not found variability in the SWP spectra, in
contrast with Feibelman (1997), who reported a variation in the flux of a factor $\sim$ 2.
 In order to discard possible reduction errors in the INES spectra, the same IUE SWP spectra available from the
MAST\footnote{\tt{https://archive.stsci.edu/iue/search.php}}
archive were retrieved. These spectra are
reduced in an independent way to that of the INES archive. No sign of flux
variability was found in MAST spectra either. Therefore, we consider that the
variability found by Feibelman (1997) should be reassessed.

\section{SYNTHETIC STELLAR SPECTRA}
\label{Section:models}

In order to carry out an accurate spectral analysis of the composite
spectrum of BD+30$^{\circ}$623, three synthetic model atmosphere grids
have been used. 

For the cool component, high-resolution spectra were synthesized in
the range 3000--8000\,$\AA$ using the suite of programs {\sc synthe}
(Kurucz 1993; Castelli \& Kurucz 2003).

For the ultraviolet range (1150--3200 \AA), low-resolution models from
the {\sc atlas9} grid of model
atmospheres\footnote{\tt{http://user.oat.ts.astro.it/castelli/grids.html}}
by Castelli \& Kurucz (2003) were used. Finally,
  each ultraviolet spectrum was combined with the corresponding
  optical one to build a grid in the range of 1150-8000 \AA.
 
For the hot component, we used TMAW\footnote{\tt{http://astro.uni-tuebingen.de/$\sim$TMAW/}}, 
 a service recently developed in the framework of the German Astrophysical Virtual Observatory
(GAVO), to calculate theoretical stellar spectra of hot, compact stars.
The service provides access to TMAP\footnote{\tt{http://astro.uni-tuebingen.de/$\sim$TMAP/}}
 (T\"{u}bingen NLTE - Model Atmosphere Package; Werner et al. 2003), a collection of models
  successfully used for spectral analysis of
hot, compact stars (e.g, Rauch et al 2007, 2013; Ziegler et al. 2012).

We requested for a H+He, high-resolution
grid, via TMAW, with 50000\,K $\le$ $T_{\rm eff}$ $\le$ 120000\,K in
the range 1000 -- 55000\,$\AA$. The spectral resolution was 0.1\,$\AA$
in the range 1000 -- 8000 \,$\AA$ and 1\,$\AA$ in the rest of the
wavelengths. Since the H and He lines are not very
sensitive to $T_{\rm eff}$ beyond 70000\,K, we sampled from 50000\,K
to 70000\,K in steps of 2000\,K and from 70000\,K to 120000\,K in
steps of 5000\,K. Surface gravity in this grid ranges from 3.0 $\le$
$\log g$ $\le$ 7.0 in steps of 0.2 dex. However we note that those
models above the Eddington limit could not be
calculated with the TMAW procedure. Therefore, the lower limit in
$\log g$ is in fact set by the Eddington limit. The abundance ratio
(by mass) of hydrogen to helium was varied as follows:
H/He = 1/0, 0.9/0.1, 0.8/0.2, ..., 0.1/0.9, 0/1. No other elements
have been taken into account for the calculation.

Before starting the spectral analysis, the synthetic spectra were
degraded and rebinned to the resolution of the observations, which is $\simeq$
2\,$\AA$\,pixel$^{-1}$.

\section{SPECTRAL ANALYSIS}
\label{Section:analysis}

\subsection{Composite models}
\label{Section:composite}
 
The observed flux of a composite spectrum can be reproduced by the sum
of two terms, each one referring to each individual spectrum,
following the relation:

\begin{eqnarray}
   F_\lambda & = & \left ( \frac{R_{\rm cool}}{d}\right)^2 S_\lambda(T_{\rm eff
     (cool)}, g_{\rm cool}) + \left (\frac{R_{\rm hot}}{d}\right)^2
   S_\lambda(T_{\rm eff (hot)}, g_{\rm hot}) \nonumber \\ 
   		     & = & k_{\rm norm} [S_{\lambda}(T_{\rm eff (cool)},g_{\rm
  cool})\,+\,a\,S_{\lambda}(T_{\rm eff (hot)},g_{\rm hot})]\label{eq:1}
\label{Eq:Flambda}
\end{eqnarray}

\noindent where $R_{\rm cool}$, $T_{\rm eff (cool)}$, $g_{\rm cool}$, $R_{\rm hot}$, 
$T_{\rm eff (hot)}$, $g_{\rm hot}$ are the radius, the $T_{\rm eff}$ and the surface gravity of each object, $d$
is the distance to the system (assuming that  both objects are at the same
  distance) and $S_\lambda$ the
  monochromatic surface flux emission of each component. In this way the parameter $a$ is defined
as $(R_{\rm hot}/R_{\rm cool})^2$ and $k_{\rm norm}$ is the normalization constant defined as $(R_{\rm cool}/d)^2$.
Each composite model will be normalized to the
observed spectrum following the procedure described by Bertone et
al. (2004). 

\begin{table}
 \centering
  \caption{IUE spectra used for the spectral analysis. The LWP spectra beyond
 $\sim$ 2400 \AA{} could not be used for the
spectral analysis (see the text).}
  \begin{tabular}{@{}cc@{}}
  \hline    
 Spectrum & Date (dd/mm/yyyy)\\
  \hline \hline
 SWP01313 + LWR01279 & 04/04/1978   \\
 SWP03813 + LWR03394 & 05/01/1979  \\
 SWP08024 + LWR06984 & 23/02/1980 \\
 SWP18042 + LWR14219 & 20/09/1982 \\
 SWP35671 + LWR15127 & 04/03/1989 \\
\hline
\end{tabular}
\label{Table:IUEspectra}
\end{table}

The last term of equation (1) has five unknowns, namely both $T_{\rm eff}$ and gravities, and the parameter $a$.
 In addition, the extinction affecting the observed
spectrum --both $R_V$ and $E(B\!-\!V)$-- is also unknown. This amounts
to a total of seven unknowns, which poses a difficult problem to be
tackled. Therefore, to arrive the final and sensible solution(s) we have
devised a method to explore the whole range of parameters by building
a grid of composite models from the individual grids, and by applying
a suite of filters that allows us to select, among all possible
combinations, those matching the observed spectral features, both at
an overall level (the shape of the spectrum, the position of the
Balmer jump) and lines (equivalent width of some absorption lines).
The outcome of the analysis will be the $T_{\rm eff}$, gravities and the ratio between the radii of the
components.

In what follows we will explain in detail the fitting procedure, until
the final solutions are reached. Fig. \ref{Figure:flowchart} shows a
flowchart describing the whole process. Note that an
  iterative process is needed because each cycle hinges on the
  parameters of the cool component that must be revised once each
  iteration is completed, until convergence is reached.

\subsection{Parameters of the cool component}
\label{Section:cool}

\begin{figure}
\includegraphics[width=0.45\textwidth]{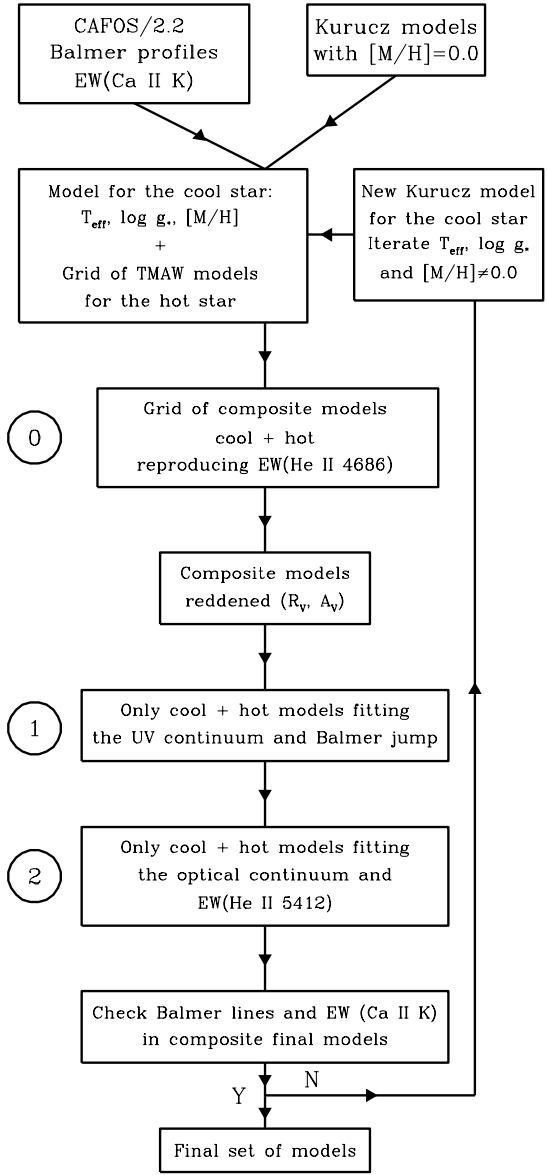}
  \vspace*{1pt}
  \caption{Flow chart with all the steps of the fitting procedure.
    See text for details.}
\label{Figure:flowchart}
\end{figure}

As it can be seen in Fig. \ref{Figure:flowchart}, and is
  described in detail in the next subsections, the whole fitting
  process is based on the computation of grids of composite models
  where, in each one, the Kurucz synthetic spectrum of the cool star
  remains fixed and it is combined with all the TMAP models. The
  composite models undergo several filters imposed by the observations
  and only the small subset that fulfills all the observational
  constraints is accepted as a final solution for the problem.
  Therefore, the first step is to constrain the parameters of the
  cool star. These parameters are reassessed at the end of each cycle.
  We note that an approach from the 
  point of view of the hot star (i.e, fixing the parameters of the 
  hot star and combining with the whole grid of the Kurucz models) is not possible since we do not 
  have enough spectral criteria to adjust the corresponding pairs of $T_{\rm eff}$ and gravity
  for the hot component.

The Ca {\sc ii} K $\lambda 3933$ line is generally one of the prime
temperature criteria in A- or B-type stars (Gray \&
  Corbally, 2009). In the case of single
stars, an estimate of $T_{\rm eff}$ can be obtained by comparing the
equivalent width (EW) of the line in the observed spectrum with
that measured on a grid of synthetic
spectra or templates. However, this simple procedure cannot be applied
in this case, because superimposed to the spectrum of the cool
component, there is an unknown contribution of the
  continuum and lines of the hot component. The only information we can
extract in the first step is an upper limit to the $T_{\rm eff (cool)}$, 
since composite models --or
combinations of templates-- containing a cool star whose EW(Ca {\sc
  ii}) $\leq$ 0.34 \AA{} (which is the EW of the line in the observed
spectrum) will have the EW of this line decreased below
that value when the spectrum of the hot companion is
added. An inspection of the grid of synthetic models (see Sect. \ref{Section:models})
 with solar metallicity sets a rough upper limit of $T_{\rm eff (cool)}$ $\sim\!11000$ K.
  The Ca {\sc ii} K line dependence on gravity and metallicity was found negligible.

Since the wings of the Balmer lines are very sensitive to gravity in
A- and B-type stars (Gray \& Corbally, 2009), we have
explored which value of $\log g$ for the cool component is consistent
with each potential $T_{\rm eff}$. We have computed synthetic
models for H$\alpha$, H$\beta$, H$\gamma$ and H$\delta$ for $T_{\rm
  eff (cool)}$=9000 -- 11500 K, in steps of 500 K, and values of $\log
g_{\rm cool}$=2.50 -- 4.00, in steps of 0.25 dex. Such intervals were chosen 
according to the typical parameters of A- and B-type stars.
The widths of the lines in the spectrum have been
compared with those in the synthetic models, and the pairs ($T_{\rm eff
  (cool)}$, $\log g_{\rm cool}$) that are consistent with the
observations are (9000 K, 2.50), (9500 K, 2.75), (10000 K, 3.00),
(10500 K, 3.25), (11000 K, 3.50) and (11500 K, 3.75). Typical errors
in the determination of $\log g_{\rm cool}$ are $\pm 0.25$ dex.

We would like to stress that we are comparing models computed with a
single $T_{\rm eff}$ and gravity with the observed spectrum, that is a
composite of two stars. The assumption underlying this exercise is
that the cool component is the main contributor to the Ca {\sc ii} K
and the Balmer lines. We also note that a possible interstellar contribution 
to this absorption line was not taken into account in the analysis. 
Although the low-resolution 
 of our spectrum does not allow us to quantify this contribution, Greenstein (1972) concluded that the cool
 component has a considerable strong stellar line by comparing the velocity 
 of the observed Ca {\sc ii} K line and the expected interstellar line velocity. 

 Once the complete filtering process is done and the contributions of
  both components to the observed spectrum are evaluated, the Ca {\sc
    ii} K line and the Balmer profiles are analyzed again and the
  values of $T_{\rm eff (cool)}$ and $\log g_{\rm cool}$ are reassessed to
  fit the observations (see Sect. \ref{Section:assessmentcool}). The
  whole fitting and filtering process is started again (see
  Fig. \ref{Figure:flowchart}).

\subsection{First filter: UV continuum and Balmer jump}
\label{Section:filter2}

Before starting with the filtering procedure, we imposed the condition that 
 the EW of the He\,{\sc ii} $\lambda$4686 line of {\it all} composite
  models matches that of the observed line, namely $\sim\!0.42$ \AA. Therefore, for each pair
  $S_{\lambda}(T_{\rm eff (cool)},g_{\rm cool})$, $S_{\lambda}(T_{\rm eff
    (hot)},g_{\rm hot})$, the parameter $a(T_{\rm eff (cool)},T_{\rm
    eff (hot)},g_{\rm cool},g_{\rm hot})$ is computed in such a way that
the synthetic EW matches the observed one. The outcome of
this step, --marked as `0' in Fig. \ref{Figure:flowchart}-- is a grid of composite models that
will be passed to the first filter.

\begin{figure*}
\includegraphics[width=1.0\textwidth]{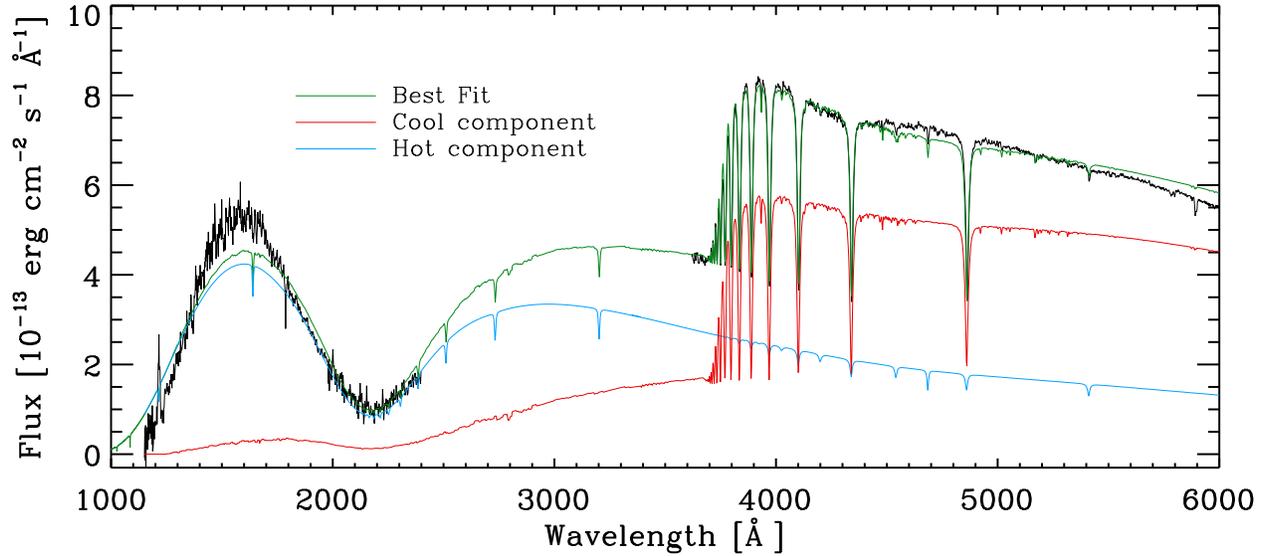}
  \vspace*{1pt}
  \caption{Observed spectrum of BD+30$^{\circ}$623 (black) and one of
    the best fits (green) chosen between the models for
      the hot component that pass the filters described in
      Sect. \ref{Section:analysis}, which corresponds to $T_{\rm
      eff (hot)}$ = 90000\,K, $\log g_{\rm hot}$ = 5.6 and H/He=
    0.4/0.6. The parameters for the cool component are
      $T_{\rm eff (cool)}$ = 9850\,K and $\log g_{\rm cool}$ = 3.5. All the
    other solutions are virtually identical and are not plotted to
    avoid confusion. In red and blue the separate
      contributions of the cool and hot stars, respectively, are plotted.}
\label{Figure:fit80000}
\end{figure*}

\begin{figure*}
\includegraphics[width=0.85\textwidth]{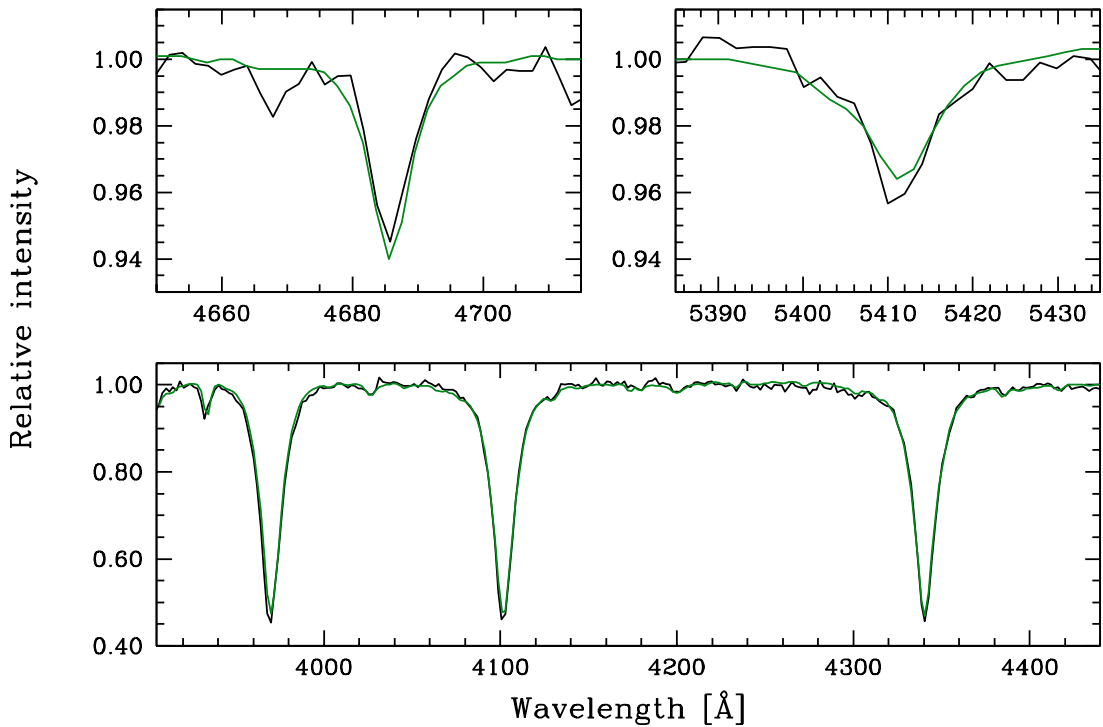}
  \vspace*{1pt}
  \caption{Profiles of the observed absorption lines (black) used for the spectral analysis together with the synthetic
  profiles (green) of one of the final solutions. In the upper part of the figure the He {\sc ii} $\lambda$4686 and He
  {\sc ii}$\lambda$5412 lines. In the lower part, the Ca {\sc ii} K line and the
  Balmer lines H$\epsilon$, H$\delta$ and H$\gamma$.}
\label{Figure:individual_lines}
\end{figure*}

Since the ultraviolet and blue wavelengths are the most sensitive to
the interstellar and nebular extinction, we carried out first an
initial analysis fitting only the UV shape and the Balmer jump. To do
this, we explored a wide range of reddening values and applied those to the
models. In principle we shall not assume the unique Galactic average value of 3.1 for $R_V$ 
(see Sect. \ref{Section:mrld} for further discussions) then, we varied this
    parameter and the colour excess $E(B\!-\!V)$. The values adopted
  for this analysis range from 1.5 to 5.0 in steps of 0.1
    in the case of $R_V$, and from 0.2 to 1.5 in steps of 0.05 in the
    case of $E(B\!-\!V)$. The parameterization of the extinction by
  Fitzpatrick (1999) has been used.

A least squares fit with a confidence level of 99.73\% (3$\sigma$),
yields a first effective filtering over the total grid. We
obtained that only those models with $E(B\!-\!V)$ between 0.5 and 0.6
and $R_V$ between 2.3 and 2.6 fitted well the UV shape, and pass to the next filter. 
This step is marked as `1' in Fig. \ref{Figure:flowchart}.

\subsection{Second filter: optical continuum and He\,{\sc ii} $\lambda$5412}
\label{Section:filter3}

After the first filter, we performed a second fitting
  --`2' in Fig. \ref{Figure:flowchart}-- to match
the optical continuum. Once again, a least squares fit with a
confidence level of 3$\sigma$ was carried out. At this point, 
we also imposed the requirement of a good fitting to the He\,{\sc ii}
$\lambda$5412 absorption line.

\subsection{Reassessment of the parameters of the cool star}
\label{Section:assessmentcool}

As mentioned in Sect. \ref{Section:cool}, in the first
  iteration we repeated the fitting procedure keeping fixed all the
  pairs ($T_{\rm eff (cool)}$, $\log g_{\rm cool}$) of the cool star that
  seemed to be compatible with the observations, combining them with
  the whole grid of synthetic spectra for the hot star.

  After that first iteration, the EW of the Ca {\sc ii} K line and the
  widths of the wings of the Balmer lines were measured on each of the 
  composite spectra that passed all the filters. As expected,
  the EW of the Ca {\sc ii} line is less, and the wings of the
  Balmer lines were narrower, than those in the observed spectrum
  --and on each of the individual models with ($T_{\rm eff (cool)}$, $\log
  g_{\rm cool}$)-- due to the fact that a contribution of the hot
  component had been added to each cool model. The comparison of the
  features of the composite spectra with those observed helped to tune
  the parameters of the cool star. A new iteration was then started with
  the revised set of parameters, as shown in Fig. \ref{Figure:flowchart}.

After a few cycles it turns out that convergence in the
  procedure, and an excellent agreement with the observed spectrum,
  are achieved for $T_{\rm eff (cool)}$ = $9850\pm 150$\,K and $\log g_{\rm
    cool}$ = $3.50\pm0.25$. These parameters are compatible with both an A0 star 
    ascending the giant branch (as reported by, e.g, Kohoutek 1967) 
    and a Horizontal-Branch (HB) A0 star (as reported by Greenstein 1972).
    We will see below that the HB A0 classification appears to be the most probable one.
    The strength of the Ca {\sc ii} K and
  other weak blends point to a subsolar metallicity. A value of
  [M/H]$_{\rm cool}\!=\!-0.5$, which seems to match the observed
  features, has been adopted; however, high-resolution spectra would
  be needed to confirm or reject this.

\subsection{Final results for the hot star}
\label{Section:finalhot}

  For that particular set of parameters of the cool star, seven
  successful solutions were obtained for the hot companion, with
  $T_{\rm eff (hot)}$ between 80000\,K and 95000\,K, $\log g_{\rm hot}$
  between 5.4 and 5.6 and compositions compatible with a helium-rich
  star (H/He= 0.0/1.0, 0.2/0.8, 0.3/0.7, 0.4/0.6, 0.5/0.5, 0.6/0.4). These final solutions
  together with the corresponding values of the parameter $a$ (see
  Eq. \ref{Eq:Flambda}) are listed
  in Table\,\ref{Table:hotstarparams}. The derived stellar parameters are 
  compatible with most of the different
known types of CSPNe, so we would need more information (e.g., mass,
radius, luminosity) to classify the hot component in a
definitely way (see Sect. \ref{Section:mrld}).

Fig. \ref{Figure:fit80000} shows one of these final
  solutions, namely that with $T_{\rm eff (hot)}$ = 90000, $\log g_{\rm
    hot}$ = 5.6 and H/He= 0.4/0.6, plotted in green. All the other solutions 
    are virtually identical and are
  not included in the figure to avoid confusion. We have also plotted
  the individual contributions of the cool and hot stars in red and
  blue, respectively. As mentioned in Sect. \ref{Section:observations}, 
  the neutral helium lines (e.g., He\,{\sc i} $\lambda$4471) come from the cool companion. 
   This is in agreement with the high $T_{\rm eff}$ derived for the hot star.

\begin{table}
 \centering
  \caption{Spectroscopic solutions for the hot component of
    BD+30$^{\circ}$623.}
  \begin{tabular}{@{}cccc@{}}   
   \hline
$T_{\rm eff (hot)}$ [K]  &  $\log g_{\rm hot}$ & H/He & $a$\\
  \hline \hline
80000 &  5.4  &   0.6/0.4  &      0.01445  \\
85000 &  5.6  &   0.5/0.5  &      0.01328  \\
90000 &  5.6  &   0.4/0.6  &      0.01211  \\
95000 &  5.6  &   0.0/1.0  &      0.01172  \\
95000 &  5.6  &   0.2/0.8  &      0.01113  \\
95000 &  5.6  &   0.3/0.7  &      0.01113  \\
95000 &  5.6  &   0.4/0.6  &      0.01094  \\
\hline
\end{tabular}
\label{Table:hotstarparams}
\end{table}

   In Fig. \ref{Figure:individual_lines} 
  the profiles of the He {\sc ii} $\lambda$4686, He
  {\sc ii}$\lambda$5412, Balmer lines H$\epsilon$, H$\delta$ and H$\gamma$, and Ca {\sc ii} K line,
  together with the corresponding fits
  are shown.

\subsection{Extinction}
\label{Section:extinction}

As it can be seen in Fig. \ref{Figure:flowchart} and
  Sect. \ref{Section:filter2}, each composite model is reddened with a
  range of values of $R_V$ and colour excess $E(B\!-\!V)$ to match the
  shape of the UV continuum and the intensity of the Balmer jump. All
  final models matching successfully the observed spectrum had to be
  reddened with values of $R_V=2.3\pm0.1$ and
  $E(B\!-\!V)=0.60\pm0.05$, which imply $A_V\!\simeq\!1.4$. Note that
  the value of the extinction does not have any impact on those
  filters matching the EW of lines, since on narrow regions around
  absorption lines the extinction correction can be considered as
  constant and its effect cancels out when computing EWs. 

The resulting value of  $\simeq$ 1.4 for $A_V$ is in good agreement with the value of 1.6 
obtained by Ressler et al. (2010). Our estimations are also
  consistent with the colour excess obtained from the nebular
  extinction coefficient derived from the CAFOS spectrum ($E(B\!-\!V)$= 0.66, 
  see Sect. \ref{Section:observations}). We note that the colour excess derived from the
  best fit is slightly higher than the previous values proposed in the
  literature, which ranges from 0.44 to 0.5 (Seaton 1980; Feibelman
  1997). 

 We also note that while the bump at 2200 \AA{} is successfully reproduced, 
the fitting around $\sim$1600 \AA{} does not reach the intensity levels of the observed spectrum. 
This can be a consequence of the parameterization used, since only a value of $R_V$ has been taken into account.
 Besides, the lack of an accurate knowledge about the behavior of the extinction laws in this spectral range
 makes it difficult this kind of analysis.

\begin{table}
 \centering
  \caption{Stellar parameters of the two components of
    BD+30$^{\circ}$623.}
  \begin{tabular}{@{}rrr@{}}
 \hline
 Stellar parameters  & Cool Star & Hot star\\
  \hline \hline
$T_{\rm eff}$ [K]        & 9850 $\pm$ 150  & $90000 \pm 10000$ \\
$\log g$                & 3.5$\pm$ 0.25   & 5.5$\pm$ 0.1      \\
$M/{\rm M}_{\odot}$   & 		0.55 $\pm$ 0.02  & $0.56 \pm 0.03$ \\
$R/{\rm R}_{\odot}$                & 2.1$\pm$ 0.6   & 0.22$\pm$ 0.03      \\
$\log(L/{\rm L}_{\odot})$      & 1.60 $\pm$ 0.25  & $3.4 \pm 0.2$ \\
$R_V$                   & \multicolumn{2}{c}{2.3 $\pm$ 0.1}   \\
$E(B\!-\!V)$            & \multicolumn{2}{c}{0.60 $\pm$ 0.05} \\
$d$ [pc]           & \multicolumn{2}{c}{285 $\pm$ 85} \\
\hline
\end{tabular}
\label{Table:twostarsparams}
\end{table}

\section{DISCUSSION}
\label{Section:mrld}

The spectral analysis described above provides
  values of the $T_{\rm eff}$ and gravities of both
  components of BD+30$^{\circ}$623. While it is true that the $T_{\rm eff (cool)}$ has been already 
  determined in a quite accurately way by several authors (see above), this is the first time that the parameters
   of the hot component have been spectroscopically derived.
    In Table \ref{Table:twostarsparams}
    we list the results derived from the spectral analysis described in the previous 
    sections ($T_{\rm eff}$ and gravities of the two components, and extinction)
     as well as the stellar parameters derived below from the evolutionary tracks 
    (mass, radius, luminosity of the two components and the distance to the system). 
    Note that the $T_{\rm eff (hot)}$
    is not an strict average of the 
    $T_{\rm eff}$ listed in Table \ref{Table:hotstarparams}, we prefer
    to give a round number that represents all the solutions with an
    uncertainty bracketing all of them. Also note that the error in  $\log g_{\rm hot}$ comes 
    from an average of the different values of the  $\log g$ 
 of the models that have passed all the filters in the analysis. In what follows, we adopt between the seven 
   successful solutions 
   matching the observed spectrum, that shown in Fig.\,\ref{Figure:fit80000}, namely, $T_{\rm eff (hot)}$ =
  90000\,K, $\log g_{\rm hot}$ = 5.6, $T_{\rm teff (cool)}$ = 9850\,K and
  $\log g_{\rm cool}$ =3.5. These parameters allow us to compare with the
  appropriate evolutionary tracks and isochrones.

\begin{figure}
\includegraphics[width=0.47\textwidth]{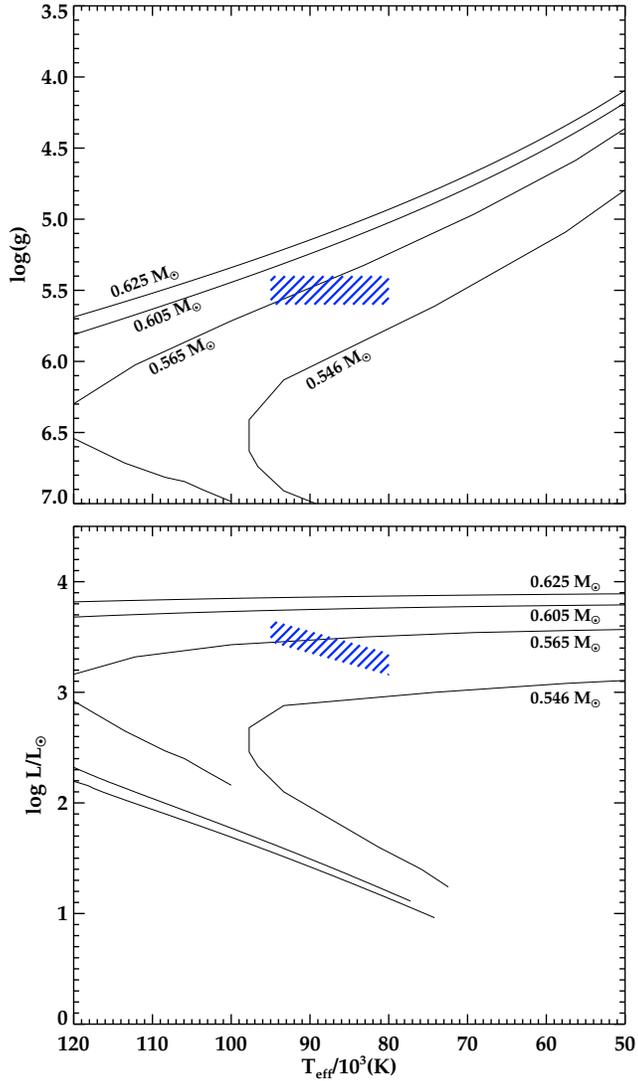}
  \vspace*{1pt}
  \caption{Location of the hot component on the evolutionary tracks by
    Bl\"ocker (1995). The blue solid lines indicates the region of the possible parameters
    of the hot star derived from the final solutions. 
    Each evolutionary track is labelled with the
    corresponding stellar mass (in ${\rm M}_{\odot}$). }
\label{Figure:HRdiagramshot}
\end{figure}

\begin{figure}
\includegraphics[width=0.47\textwidth]{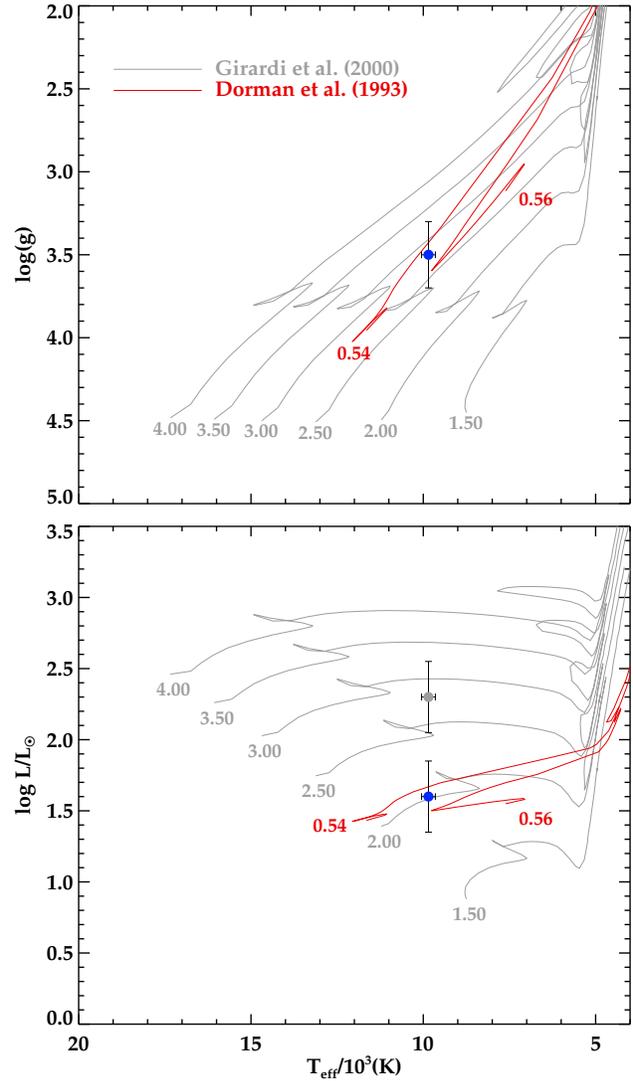}
  \vspace*{1pt}
  \caption{Location of the cool component on the evolutionary
    tracks calculated by Girardi et al. (2000) (grey lines) and the post-HB evolutionary tracks by 
    Dorman et al. (1993) (red lines). Each evolutionary track is labelled with the
    corresponding stellar mass (in ${\rm M}_{\odot}$). The location of the cool component in the top panel
   corresponds to two positions in the bottom panel depending on the evolutionary 
   track with which we are comparing it (see the text).}
\label{Figure:HRdiagramscool}
\end{figure}

Fig. \ref{Figure:HRdiagramshot} shows the position of the hot component
  of BD+30$^{\circ}$623 in the HR diagrams $\log g_* -
  T_{\rm eff}$ (top) and $\log (L_*/{\rm L}_{\odot}) - T_{\rm eff}$
  (bottom) on the evolutionary tracks for post-AGB tracks by Bl\"ocker (1995).
  The location of the star is consistent with a mass of
    $M_{\rm hot}\!=\!0.56\pm 0.03$ ${\rm M}_{\odot}$, which implies a
  radius  $R_{\rm hot}\!=\!0.22\pm 0.03$ ${\rm
      R}_{\odot}$, and a luminosity $\log(L_{\rm
      hot}/{\rm L}_{\odot})\!=\!3.4\pm 0.2$. The luminosity and the dereddened integrated flux of
the model fitting this component lead to a distance $d_{\rm hot}$=253$\pm$88 pc. 
    The derived stellar mass indicates a progenitor of $\sim$ 1 ${\rm M}_{\odot}$ in the main sequence
   and, therefore, an age of about $10^{9}$-- $10^{10}$ yr.
  The parameters derived from the evolutionary tracks agree with those typical of 
  sdOs (Oreiro et al. 2004; Geiger et al. 2011). 
 Also, the narrowness of the He\,{\sc ii} $\lambda$4686 absorption
line is compatible with an sdO classification (as previously 
   proposed by e.g. Kohoutek 1967, Greenstein 1972), rather than with a white dwarf. 
 However, other classifications (like, e.g, O(He)-type) may also be compatible. 
 Therefore, an analysis of high-resolution spectra in the UV range, as
that carried out by Latour et al. (2013) for the sdO BD+28\,4211, could
allow us to confirm or rule out a possible sdO classification.

  Fig. \ref{Figure:HRdiagramscool} shows the position of the cool
  component in the HR diagrams $\log g_* - T_{\rm eff}$ (top) and
  $\log (L_*/{\rm L}_{\odot}) - T_{\rm eff}$ (bottom) on the post
  main-sequence evolutionary tracks by Girardi et al. (2000) with
  $Z\!=\!0.004$ (grey) and the post-HB evolutionary tracks by Dorman
  et al. (1993) with $Z\!=\!0.006$ (red). The stellar parameters
  derived are compatible, a priori, with two possible scenarios: an A0
  star ascending the giant branch and a A0 star in the HB.

  In the hypothesis that the cool component were an A0 star ascending the
giant brach, the comparison with the Girardi et al. (2000) tracks gives a
mass $M_{\rm cool}\!=\! 2.9 \pm 0.5$ ${\rm M}_{\odot}$, which implies a
radius $R_{\rm cool}\!=\! 4.9 \pm 1.5$ ${\rm R}_{\odot }$, a luminosity
$\log(L_{\rm cool}/{\rm L}_{\odot})\!=\!2.3\pm 0.3$ and an age $\sim\!300$
Myr. These parameters lead to a number of inconsistencies when compared
with those of the hot star. If the CS is a physical pair, the age of the
cool star is incompatible with the much longer time span required by the
hot star to reach its current evolutionary status. Therefore, interaction
between the stars (e.g., mass transfer) should be invoked to explain the
age difference. This inconsistency in the age has been already reported for 
other similar CSs (see e.g., NGC 3132, Ciardullo et al. 1999). However, the luminosity and the
dereddened integrated flux of the cool star lead to a distance $d_{\rm
cool}$=710$\pm$280 pc that is incompatible with the CS being a physical binary but a
coincidental alignment. This scenario is also difficult to reconcile with
the observations. In particular, it would be hard to explain that the
Hubble Space Telescope does not resolve the pair, {\it and, simultaneously,} that the mean radial velocity of the
cool star ($V_{\rm LSR}$ = +47.6$\pm$1.6\,km\,s$^{-1}$) and the systemic
velocity of the nebula ($V_{\rm LSR}$ = +41$\pm$5\,km\,s$^{-1}$) are in
agreement (Greenstein 1972). In addition, the complex nebular
morphology is difficult to understand within a single-star hypothesis but
strongly suggests the presence of a binary system (see below). Moreover,
the radii derived from the evolutionary tracks yield a value of 0.002 for
the parameter $a$, defined as $(R_{\rm hot}/R_{\rm cool})^2$ under the
assumption that both components form a binary system (eq.
\ref{Eq:Flambda}); that value is a factor of 6.0 lower than the mean value
$a$=0.012 obtained from the spectral analysis
(Table\,\ref{Table:hotstarparams}). In view of all these arguments we can
discard the scenario of the cool component being a giant-branch star.
  
  Alternatively, in the hypothesis that the cool component were an A0 star in the HB, 
  the comparison with the Dorman et al. (1993) tracks gives a mass
  $M_{\rm cool}\!=\! 0.55
    \pm 0.02$ ${\rm M}_{\odot}$  (Fig. \ref{Figure:HRdiagramscool}, red lines),
    $R_{\rm cool}\!=\! 2.1 \pm 0.6$ ${\rm R}_{\odot }$ and $\log(L_{\rm cool}/{\rm
      L}_{\odot})\!=\!1.60\pm 0.25$. 
      The luminosity and the dereddened integrated flux lead to a distance
$d_{\rm cool}$=294$\pm$69 pc that agrees very well with that derived above for the hot component 
($d_{\rm hot}$=253$\pm$88 pc) and is compatible with both stars forming a binary system. 
Moreover, in contrast with the giant-branch scenario, the radius derived for the HB star 
gives $a$ = 0.011, 
in very good agreement with the mean value of $a$ obtained from the spectral 
analysis (see Sect. 4). 
Besides, the normalization constant
 between the best fitting models and the
observed spectrum is $k_{\rm norm}\!\simeq\!2.77\times 10^{-20}$, which 
is defined as $(R_{\rm cool}/d)^2$ (eq. 
\ref{Eq:Flambda}). Taking $R_{\rm cool}\!=\! 2.1$ we obtain a distance to the system
d=285$\pm$85 pc, in agreement with the values derived independently for each component. 
This result also agrees with the range 200-300 pc considered by Ressler
et al. (2010) based on the Hipparcos parallax (185 pc) and, alternatively, 
on the observed $m_V$ magnitude and the assumption 
that the cool star is an HB A star with M$_{V}\simeq$ 1.0 (240 pc). 
The fact that we obtain a very similar distance from an independent way --in our case, a spectral analysis--
 strongly reinforces the validity of our results. Finally, it should be noted that
the parallax from the Hipparcos catalogue is $\pi\!=\!5.40\pm 1.70$ mas
and the new reduction of the Hipparcos astrometric data by van Leeuwen
(2007) provides $\pi\!=\!3.79\pm 1.61$ mas. In both cases
$\sigma_\pi/\pi>0.17$ and the Lutz-Kelker bias (Lutz \& Kelker 1973)
prevents from using the Hipparcos parallax to obtain a reliable distance.

Concerning the age of the cool component, we note that the Dorman et al. (1993) evolutionary tracks do
 not provide neither
 the masses of the progenitor stars in the main sequence (ZAMS) nor their current ages. 
 However, according to the evolutionary models by Serenelly \& Weiss (2005), stars with masses 
 between 0.8 and 0.9 ${\rm M}_{\odot}$ in the ZAMS might evolve 
 (depending on the metallicity and the mass loss) to HB stars with masses 
 around 0.55 ${\rm M}_{\odot}$ --as it is our case-- 
 in periods between 8 and 12 Gyr. This range of ages is comparable
  to the age estimated for the hot component of BD+30$^{\circ}$623. Therefore, the classification 
  of the cool component as an HB A0 star is completely coherent 
with both the spectral analysis and the subsequent evolutionary study based on tracks and isochrones.

Another interesting point is the complex morphology of NGC\,1514,
consisting of numerous bubbles in the optical range (see Fig.\,1)
and a pair of axisymmetric rings recently discovered at mid-infrared
wavelengths (Ressler et al. 2010). As mentioned above, a binary nature of
BD+30$^{\circ}$623 could already be tentatively inferred from this complex
morphology, since binary interactions are generally believed to be the
origin of non-spherical structures in PNe (see, e.g., Miszalski et al.
2009; de Marco 2009). It is true that neither radial velocity variations
nor photometric and spectroscopic variability have been found in
BD+30$^{\circ}$623 (Greenstein 1972; Purgathofer \& Schnell 1983, see also
Sect. \ref{Section:IUE_spectra}), which does not favor an association.
However, in our case, the binary could be relatively close or have evolved
through a common envelope phase (see Muthu 
\& Anandarao 2003) but being observed pole-on (or nearly) so that no variability
can be observed whereas binary interactions can shape the complex nebula.
We note that a pole-on binary is compatible with the narrowness of the
absorption lines in the A0 star reported by Greenstein (1972), who
concluded that the rotational velocity has to be small ($\leq$
40\,km\,s$^{-1}$). Alternatively, the pair could be a wide binary. This
scenario would explain the lack of variability but have difficulties to account
for the nebular complexity, unless particular orbital properties are
considered (e.g., a highly eccentric orbit). In any case, conclusions
about the characteristics of the binary and its relationship to nebular
shaping should await for higher-resolution images of this object.

NGC 1514 is not unique among PNe and cases of
PNe with ``cool" CSs are well known (Lutz 1979, de Marco 2009).
In particular, He\,2-36, NGC 2346, NGC 3132, and PC\,11 (M\'endez 1978; Pereira et
al. 2010) contain an A-type star like NGC 1514.
Moreover, most of these PNe present bipolar
or complex morphologies that are much easier to explain in a binary
scenario than in a single star one. 
We note that the knowledge of the stellar parameters
 for the hot components in these peculiar CSPNe is very scarce and the analysis of 
 these CSs has been mainly focused on the cool components (see Table\,4 in de Marco 2009). 
 In this context, the detailed spectral
  analysis presented in this work introduces a remarkable improvement since it allows us to obtain the stellar
   parameters of both cool and hot component simultaneously. Besides, this kind of analysis is useful to 
   refine the spectral classification and the evolutionary status of the cool components in these systems.
The future application of this method to similar peculiar CSs will be highly interesting 
to search for common properties in these systems and to constrain the formation 
and evolution of these enigmatic pairs of stars.

\section{CONCLUSIONS}
\label{Section:conclusions}

We have analysed ultraviolet IUE and CAFOS optical spectra of
BD+30$^{\circ}$623, the peculiar binary central star of NGC 1514. BD+30$^{\circ}$623 shows an unusual
  central star spectrum whose main features correspond to an A-type star but
  also presents He {\sc ii} lines that must originate in a much hotter
  companion. Previous studies of this central star have
  been unable to provide unambiguous parameters for its two components. 
  Making use of state-of-the-art model-atmospheres for hot, compact
stars (T\"{u}bingen NLTE - Model Atmosphere Package) and Kurucz
synthetic models, we have carried out a detailed analysis of the
ultraviolet and optical spectra with the aim of constraining the
properties of this object. Grids of composite models of a cool+hot star were built. 
A $\chi^{2}$-fit and an iterative
procedure based on filters imposed by observational constraints was
devised to find the solutions that best reproduce both the shape of
the ultraviolet and optical continuum and the main spectral features. To our knowledge, this kind of detailed analysis 
has not been done before for any of the peculiar binary central stars.
The main results of this work can be summarized as follows:

1. The spectroscopic analysis leads to $T_{\rm eff
    (cool)}\!=\!9850\pm 150$ K, $\log g_{\rm cool}\!=\!3.50\pm 0.25$ and
  [M/H]$\simeq\!-0.5$.  For the
hot component, solutions for $T_{\rm eff (hot)}$ between 80000\,K and 95000\,K,
$\log g_{\rm hot}$ between 5.4 and 5.6 and compositions pointing to a
helium-rich star were found.
An unique value for the extinction towards the two stars has been assumed, namely, 
$R_V\!=\!2.3\pm0.1$ and $E(B\!-\!V)\!=\!0.60\pm0.05$. Besides, the extinction coefficient derived
 from the nebular spectrum is $c$(H$\beta$) $\simeq$ 0.97.

 2. By comparing the stellar parameters  derived for both components with the appropriate 
  evolutionary tracks, we conclude that the cool component is compatible with an Horizontal-Branch
  A0 star with mass $M_{\rm cool}\!=\! 0.55
    \pm 0.02$ ${\rm M}_{\odot}$, radius 
    $R_{\rm cool}\!=\! 2.1 \pm 0.6$ ${\rm R}_{\odot }$ and luminosity $\log(L_{\rm cool}/{\rm
      L}_{\odot})\!=\!1.60\pm 0.25$. A distance $d_{\rm cool}$=294$\pm$69 pc is also derived. 
      In the hypothesis underlying all this work that the stars form a physical pair, the scenario 
      with an A0 star ascending the giant branch proposed by other authors seems to be 
      inconsistent and therefore, should be 
      discarded. On the other hand, the parameters of the hot component 
      are $M_{\rm hot}\!=\!0.56\pm 0.03$ ${\rm M}_{\odot}$, $R_{\rm hot}\!=\!0.22\pm 0.03$ ${\rm
      R}_{\odot}$, and $\log(L_{\rm cool}/{\rm L}_{\odot})\!=\!3.4\pm 0.2$. A distance of $d_{\rm hot}$=253$\pm$88 pc 
      is derived. Although an sdO nature seems to be the most plausible one for this 
  component, it is difficult to establish this classification firmly without high-resolution observations. 
  The analysis is compatible with 
  the two stars forming a physical pair.

 3. A reanalysis of the IUE spectra shows that no variability is seen at short wavelengths, in
contrast with the variation of the continuum flux by a factor of $\sim\!2$ reported by Feibelman (1997).

The analysis presented in this paper improves noticeably previous
studies carried out for this object. However, further observations, like 
high-resolution spectra, will
help to determine with a better accuracy parameters such as the metal
abundance of the cool component and the rotation velocities. New long- and short-term observations would be also 
required in order to study the possible variability of this system and the properties of the binary.
Theoretical studies addressing the evolution of this kind of systems are also imperative.

\section*{Acknowledgments}

We are very grateful to our anonymous referee for his/her comments that 
have improved the discussion of the paper.
This work has been partially supported by grant AYA\,2011-24052 (AA, ES), AYA2011-26202 (BM), and 
AYA\,2011-30228-C3-01 (LFM) of
the Spanish MINECO, and by grant INCITE09\,312191PR (AA, LFM, AU) of Xunta de
Galicia, all of them partially funded by FEDER funds, and by grant 12VI20 (AA, LFM, AU) of the University of Vigo. 
The authors thank Nicole Reindl for her helpful suggestions regarding the TMAP models, and Jes\'us Ma\'{\i}z-Apellaniz 
for discussions about the extinction. The TMAW service (http://astro-uni-tuebingen.de/~TMAW) used to
calculate theoretical spectra for this paper was constructed as part
of the activities of the German Astrophysical Virtual
Observatory. Authors also acknowledge the Calar Alto Observatory for
the service observations. We acknowledge support from the Faculty of the European
 Space Astronomy Centre (ESAC). This research is based on INES data from the IUE satellite. 
  The INES archive is supported by the Spanish Virtual Observatory through grant AYA2011-24052. 
  We have also made use of the SIMBAD database, operated at the CDS, Strasbourg (France), Aladin 
  and NASAÕs Astrophysics Data System Bibliographic Services.

\label{lastpage}

\end{document}